\newcommand{\CLASSINPUTbottomtextmargin}{4.16cm}

\documentclass[conference,a4paper]{IEEEtran}
\usepackage[left=1.43cm,right=1.43cm,top=1.8cm,bottom=4.16cm]{geometry}
\IEEEoverridecommandlockouts
\usepackage{cite}
\usepackage{amsmath,amssymb,amsfonts,bbm}
\usepackage{graphicx}
\usepackage{textcomp}
\usepackage[dvipsnames]{xcolor}
\usepackage{multirow}
\usepackage{subcaption}
\usepackage{caption}\usepackage{tikz}
\usepackage{tkz-tab}
\usetikzlibrary{automata,arrows,positioning,calc}
\usetikzlibrary{shapes,snakes}
\usetikzlibrary{arrows}
\usepackage{dsfont}
\usepackage{soul}
\usepackage{subfiles}
\usepackage{comment}
\usepackage{algorithm}
\usepackage{algpseudocode}
\usepackage{amsmath}
\usepackage{amssymb}
\usepackage{amsthm}
\usepackage{booktabs}
\usepackage{epstopdf} 

\def\BibTeX{{\rm B\kern-.05em{\sc i\kern-.025em b}\kern-.08em
    T\kern-.1667em\lower.7ex\hbox{E}\kern-.125emX}}

\newcommand{\hlc}[2][yellow]{{%
    \colorlet{foo}{#1}%
    \sethlcolor{foo}\hl{#2}}%
}

\usepackage{tikz}
\usetikzlibrary{fit,calc}

\colorlet{pink}{red!40}
\colorlet{blue}{cyan!50}

\begin{document}

\bstctlcite{IEEEexample:BSTcontrol}

\title{``It's Your Turn'': A Novel Channel Contention Mechanism for Improving Wi-Fi's Reliability}

\author{
\IEEEauthorblockN{Francesc Wilhelmi$^{\star}$, Lorenzo Galati-Giordano$^{\star}$, Gianluca Fontanesi$^{\star}$\vspace{0.1cm}
}
\IEEEauthorblockA{$^{\star}$\emph{Radio Systems Research, Nokia Bell Labs, Stuttgart, Germany}}
}

\maketitle

\thispagestyle{plain}
\pagestyle{plain}

\begin{abstract}
The next generation of Wi-Fi, i.e., the IEEE 802.11bn (aka Wi-Fi 8), is not only expected to increase its performance and provide extended capabilities but also aims to offer a reliable service. Given that one of the main sources of unreliability in IEEE 802.11 stems from the current distributed channel access, which is based on Listen-Before-Talk (LBT), the development of novel contention schemes gains importance for Wi-Fi~8 and beyond. In this paper, we propose a new channel contention mechanism, ``It's Your Turn'' (IYT), that extends the existing Distributed Coordination Function (DCF) and aims at improving the reliability of distributed LBT by providing ordered device transmissions thanks to neighboring activity awareness. Using simulation results, we show that our mechanism strives to provide reliable performance by controlling the channel access delay. We prove the versatility of IYT against different topologies, coexistence with legacy devices, and increasing network densities.
\end{abstract}


\section{Introduction}
\label{sec:introduction}

With Wi-Fi 8 standardization (IEEE 802.11bn) in full swing, the next generation of such a popular technology is expected to offer not only higher throughput ($+25\%$ MAC throughput) but also better reliability ($-25\%$ worst-case latency, $-25\%$ packet losses) and mobility support~\cite{GalGerCar2023}. To meet those requirements, Wi-Fi will need, once again, to evolve and improve its existing features and include new ones. And for that, one prominent area for improvement is the shared channel access, which stems from the particular properties of the unlicensed spectrum (free and open to anyone) used by Wi-Fi devices. 

The challenge, however, stems from the fact that unlicensed frequency bands are subject to regulation such as Listen-Before-Talk (LBT), which is realized by Wi-Fi through the Distributed Coordination Function (DCF). The DCF combines Carrier Sense Multiple Access (CSMA) with Collision Avoidance (CA) and Binary Exponential Backoff (BEB), both aiming at minimizing and mitigating the collisions originating from the decentralized spectrum access. However, the DCF is well known to suffer inefficiencies and to be one of the main sources of unreliability in Wi-Fi, especially for dense networks~\cite{abinader2014performance}. One side effect of the DCF stems from its randomized backoff, whereby high latency peaks can be experienced in the short term when unlucky selecting high backoff values, which leads to degraded worse-case latency and, hence, negatively impacts low-latency applications---e.g., Virtual Reality (VR) applications may support between $2$~ms and $5$~ms maximum delay~\cite{carrascosa2024performance}.


In this paper, we propose a new channel contention mechanism for Wi-Fi, named ``It's Your Turn'' (IYT), which builds upon legacy DCF and aims at improving reliability and offering consistent performance, thus contributing to achieving determinism in Wi-Fi, whilst preserving the decentralized principles of LBT. IYT aims at providing reliability by mitigating the significant delays caused by the randomness in computing the backoff values and, additionally, by reducing the number of collisions resulting from the concurrent channel access (i.e., which occur when two or more devices select the same random backoff). The main contributions of this paper are:
\begin{itemize}
    \item We propose IYT, a decentralized channel contention mechanism for Wi-Fi.
    \item We showcase the operation and main properties of the proposed IYT mechanism by using representative scenarios, conceived to provide a clear understanding of the behavior of the mechanism.
    \item We evaluate the proposed IYT mechanism through extensive simulations and compare its performance against two relevant baseline methods, BEB (adopted in IEEE 802.11) and Deterministic Backoff (DB)~\cite{kosek2022db}.
 \end{itemize}

 The rest of the paper is structured as follows. Section~\ref{sec:related_work} reviews the state-of-the-art concerning decentralized channel access. Section~\ref{sec:proposed_mechanism} describes the proposed IYT mechanism and the two considered baselines, BEB and DB. Simulation results are reported in Section~\ref{sec:performance_evaluation} and Section~\ref{sec:conclusions} summarizes the main findings and provides final remarks.

\section{Related Work}
\label{sec:related_work}

The core of Wi-Fi's decentralized operation is channel contention---which is based on Contention Window (CW) selection---since it determines the time in which a device can initiate a transmission in the presence of other contenders. CW optimization is a key area for providing efficient channel access and addressing the issues of wireless coexistence whilst ensuring a fair and effective usage of shared resources. However, CW optimization entails multiple challenges and is often a non-trivial task, provided that wireless networks are highly dynamic (e.g., varying interference and traffic patterns) and heterogeneous in terms of devices' capabilities and conditions.

Methods based on exponential backoff have been widely adopted in wireless networks for CW selection due to their ability to deal with collisions. BEB~\cite{barcelo2009learning}, for instance, is a well-known algorithm that mitigates collisions by increasing the CW exponentially, so that the likelihood of future collisions is reduced (more details are provided later in Section~\ref{sec:proposed_mechanism}). Other mechanisms like the Enhanced Distributed Channel Access (EDCA) have also included traffic awareness into CW adaptation so that high-priority traffic (using lower CW values) can be transmitted before non-priority traffic (using higher CW values)~\cite{tinnirello2009rethinking}. Alternatively, other CW adjustment proposals are based on the channel status~\cite{hong2012channel} or use machine learning~\cite{gawlowicz2021distributed}. 

Closer in spirit to our work, we find proposals that specifically target reliability and determinism. One example is Z-MAC~\cite{rhee2005z}, a protocol that combines CSMA and Time-Division Multiple Access (TDMA) to establish priorities over pre-defined slots. Another relevant proposal is the Deterministic Backoff (DB) mechanism defined in~\cite{kosek2022db}. DB alternates deterministic and random backoff stages, where the deterministic backoff is computed based on the number of interruptions observed during contention, thus allowing to generate certain awareness regarding nearby contending devices. DB has been presented as a potential feature for Wi-Fi at IEEE 802.11 standardization.

Centralized control (or full knowledge of the network) has also been considered for backoff management as a way to introduce determinism, thus allowing for reduced collisions and increased efficiency. An example of a centralized solution is proposed in~\cite{patras2016rigorous} for intra-BSS operation, whereby the average frame transmission duration is estimated using an analytical model and then used to compute the CW values that each device should use to maximize the proportional fairness of the network throughput. Similarly, \cite{kim2017centralized} proposes a method for handling a unique backoff state, which it uses to allocate a backoff counter to each node in a centralized manner. To realize coordinated CW management in coexisting Wi-Fi networks, we find Multi-Access Point Coordination (MAPC) for Wi-Fi 8 (expected for 2028)~\cite{GalGerCar2023}, a framework that would allow different APs to communicate and cooperate for using frequency resources more efficiently. Other coordinated mechanisms have been proposed for cellular unlicensed protocols like License Assisted Access (LAA) or New Radio Unlicensed (NR-U). In~\cite{huawei2017coexistence}, a self-deferral method for NR-U is proposed to coordinate the backoff from several gNodeB (gNB) so that they start their transmissions simultaneously. Likewise, the work in~\cite{lagen2019new} describes a coordinated LBT mechanism for NR-U whereby a gNB can share part of their ongoing transmissions to allow other coordinated gNBs to decrease their backoff.







In this paper, we propose a decentralized CW adaptation mechanism that targets high reliability by considering the estimated inter-BSS activity, which is used to determine the order at which devices must access the channel. Our proposed mechanism is fully compatible with the DCF and does not incur any additional overheads. Moreover, our mechanism allows implementing any ordering policy on top of it, thus going beyond the Round Robin (RR) type of ordering of~\cite{kosek2022db} and enabling the realization of multiple channel access priorities.

\section{Description of the Proposed Mechanism}
\label{sec:proposed_mechanism}

The access to the channel in the unlicensed spectrum is mandated by LBT, whereby a given device must determine that the channel is idle before transmitting a frame, so that it does not interfere with any other transmissions. LBT principles are embodied in CSMA/CA in the following manner. First, the channel status is determined by performing carrier sensing, whereby Signal Detect (SD) and Energy Detect (ED) conditions must be met to determine whether the channel is idle or free. Otherwise, it is detected as busy. For Wi-Fi signals, the Clear Channel Assessment (CCA) threshold is used to determine the status of the channel (any signal above the CCA would lead to a busy channel status). Then, to determine the moment at which a device can access the channel, a random backoff must be exhausted. Such a backoff can only be decreased as long as the channel is detected as idle. Otherwise, the backoff must be frozen until the channel is detected idle again.

In the following, we describe BEB, DB, and IYT, the three considered backoff computation mechanisms of this paper. The operation of the three mechanisms is summarized in 
Alg.~\ref{alg:backoff_computation}, where the unique steps of BEB, DB, and IYT are highlighted in \hlc[pink!30]{red}, \hlc[NavyBlue!30]{blue}, and \hlc[ForestGreen!30]{green}, respectively, and the rest of the steps are common for all the methods.

\algdef{SE}[SUBALG]{Indent}{EndIndent}{}{\algorithmicend\ }%
\algtext*{Indent}
\algtext*{EndIndent}
\begin{algorithm}[t!]
\footnotesize
\caption{IEEE 802.11 DCF channel access with \hlc[pink!30]{BEB}, \hlc[NavyBlue!30]{DB}, and \hlc[ForestGreen!30]{IYT} implemented by a device $j$.}\label{alg:backoff_computation}
\begin{algorithmic}[1]
    \State \textbf{Initialize:} CW$_{0}$, $n$, $N_\text{max}$, \hlc[NavyBlue!30]{$b$}, \hlc[NavyBlue!30]{$\text{IPT}$}, \hlc[ForestGreen!30]{$\text{CW}_\text{min}$}, \hlc[ForestGreen!30]{$\text{CW}_\text{max}$}, \hlc[ForestGreen!30]{$\mathcal{L}_j = \{\emptyset\}$}, \hlc[ForestGreen!30]{$T = \emptyset$}, \hlc[ForestGreen!30]{$d = 0$}
    \While {Backoff ($BO$) is active}
    \State \textbf{on} \texttt{detect-802.11-transmission-start}($i$):
    \Indent
       \If{$\mathcal{P}(j) < \text{CCA}$} 
            \State \text{CHANNEL\_STATUS} $\leftarrow$ \text{IDLE}
            \State $BO \leftarrow BO - 1$
        \Else
           \State \text{CHANNEL\_STATUS} $\leftarrow$ \text{BUSY}
           \State \hlc[NavyBlue!30]{$\text{IPT}\leftarrow \text{IPT} + 1$}
           \If{{$i \notin \mathcal{L}_j$}}
                \State \hlc[ForestGreen!30]{$\mathcal{L}_j \cup i$}
           \EndIf
        \EndIf
    \EndIndent

    \State \textbf{on} \texttt{detect-802.11-transmission-end}($i$):
        \Indent
       \If{{$i \in \mathcal{L}_j$}}
            \State \hlc[ForestGreen!30]{$T = \mathcal{L}_j\big(i+1 \mod |\mathcal{L}_j|\big)$}
            \State \hlc[ForestGreen!30]{$d \leftarrow \text{Distance}(T, j)$}
       \EndIf
    \EndIndent

    \If{BO == 0}
    \State \textproc{Start Transmission}
    \EndIf
    
    \EndWhile
    \Procedure{Start Transmission}{}
        \State Initiate a packet transmission.
           \If{Transmission is successful} 
                \State $n = 0$
           \Else
                \State $n = n + 1$
           \EndIf
           \State \textproc{Compute Backoff($n$, \hlc[NavyBlue!30]{$\text{IPT}$}, \hlc[ForestGreen!30]{$\mathcal{L}_j$}, \hlc[ForestGreen!30]{$T$})}
          \State \hlc[NavyBlue!30]{$\text{IPT} \leftarrow 0$}
    \EndProcedure
    \Procedure{Compute Backoff}{}
        \If{$n > 0$}
            \State $\text{CW} \leftarrow
            \text{CW}_0 \cdot 2^{\min(n, N_\text{max})}$  
            \State \hlc[NavyBlue!30]{$BO \sim U[0,\text{CW}-1]$}
        \Else
            \State \hlc[NavyBlue!30]{$BO \leftarrow b + \text{IPT}$}
        \EndIf     
        \State \hlc[pink!30]{$BO \sim U[0,\text{CW}-1]$}
        \State \hlc[ForestGreen!30]{$[\text{CW}_\text{min}, \text{CW}_\text{max}] \leftarrow [d \cdot \text{CW}_{0},(d+1) \cdot \text{CW}_{0}]$}        
        \State \hlc[ForestGreen!30]{$BO \sim U[\text{CW}_\text{min}-1,\text{CW}_\text{max}-1]$}        
    \EndProcedure        
\end{algorithmic}
\end{algorithm}

\begin{figure}[t!]
    \centering
    \begin{subfigure}{\columnwidth}
        \centering
        \includegraphics[width=\linewidth]{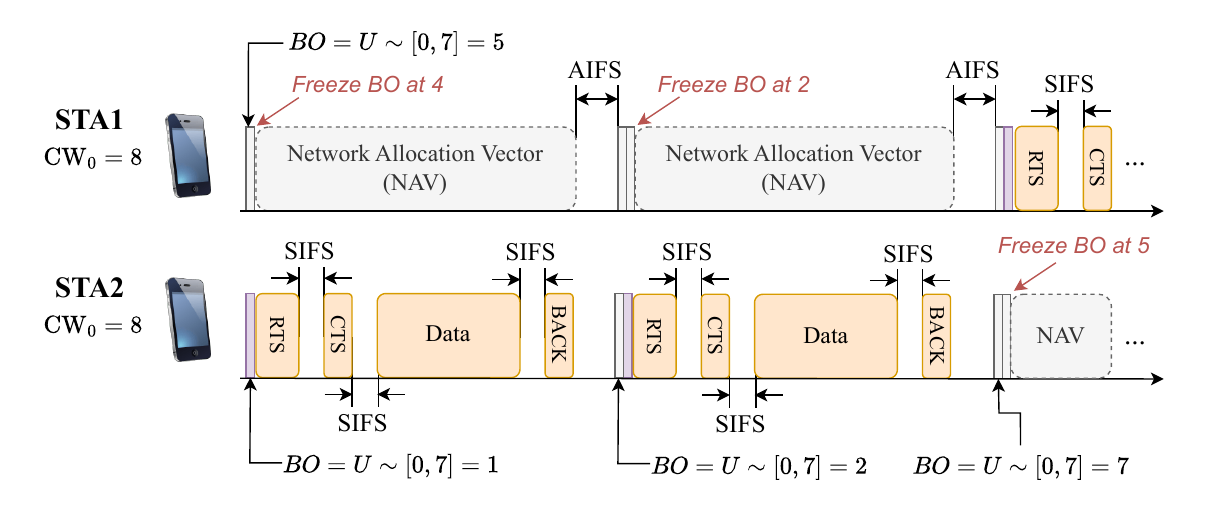}
        \caption{}
        \label{fig:beb}
     \end{subfigure}
     \hfill
     \begin{subfigure}{\columnwidth}
        \centering
        \includegraphics[width=\linewidth]{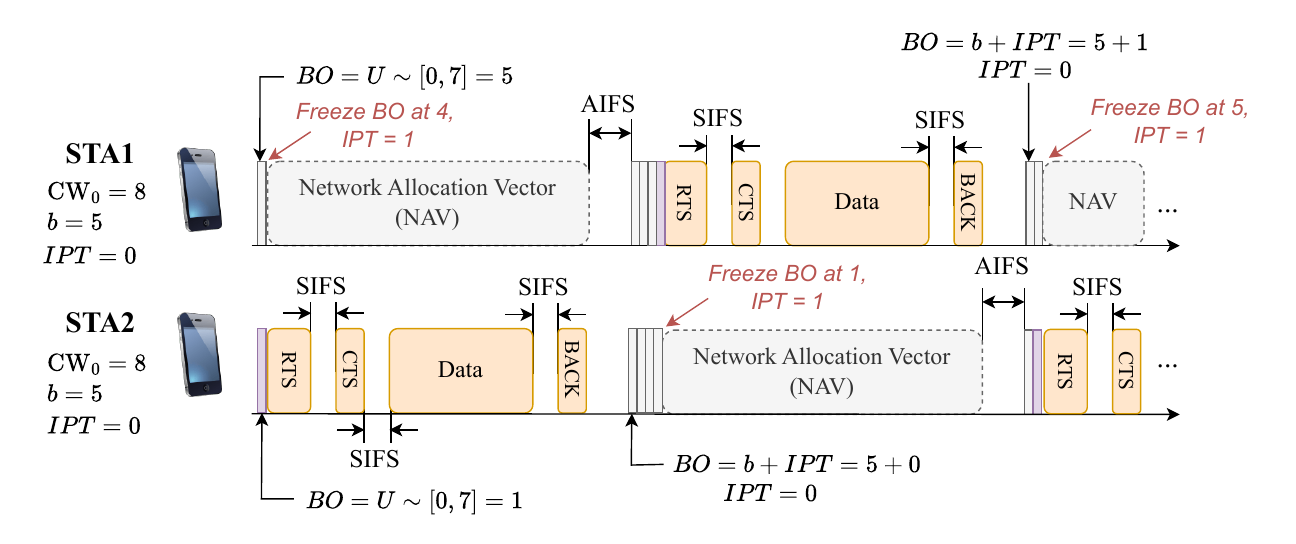}
        \caption{}
        \label{fig:db}
     \end{subfigure}
     \begin{subfigure}{\columnwidth}
        \centering
        \includegraphics[width=\linewidth]{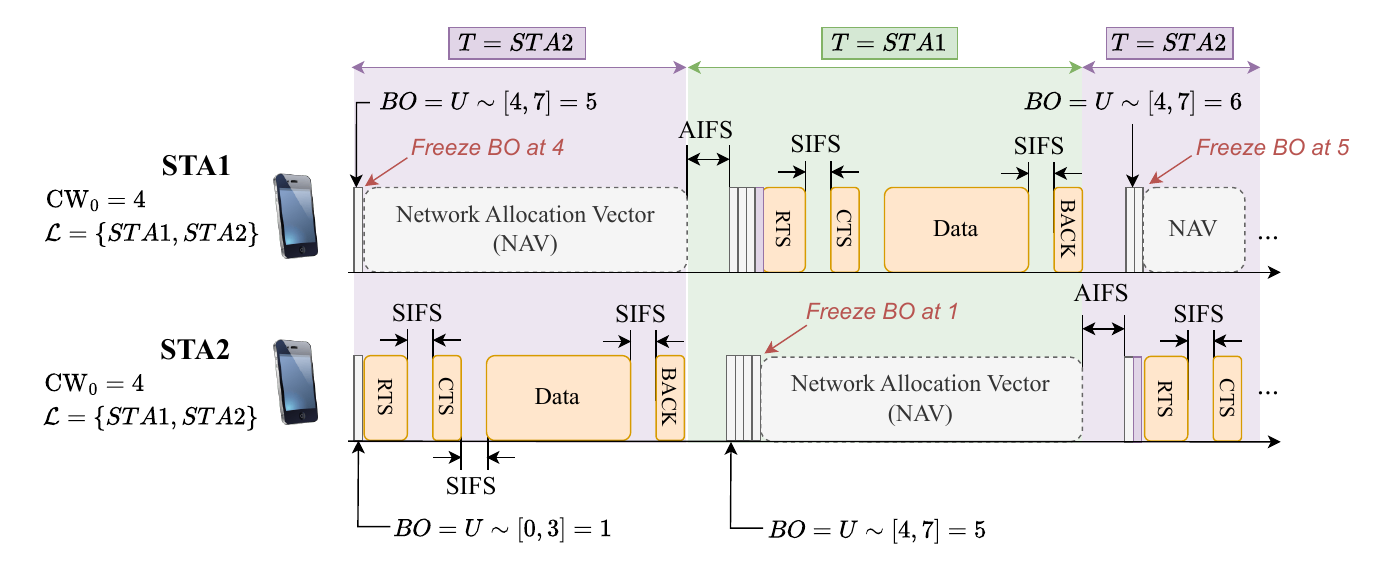}
        \caption{}
        \label{fig:iyt}
     \end{subfigure}
     \hfill
     \caption{Channel access and packet transmission diagrams for two overlapping STAs. (a) BEB (b) DB, (c) IYT.}
    \label{fig:transmission_diagrams}
\end{figure}

\addtolength{\topmargin}{+0.1cm}

\subsection{Binary Exponential Backoff (BEB)}

Exponential backoff is adopted as the default method in the IEEE 802.11 to determine the moment in which a Wi-Fi device accesses the channel. In BEB, the backoff $BO$ is selected uniformly at random between 0 and a CW parameter, i.e., $BO \sim U[0,\text{CW}-1]$ (Alg.~\ref{alg:backoff_computation}, line $40$). To mitigate collisions by backoff, which occur when two or more devices select the same random backoff, BEB adopts an exponential phase (Alg.~\ref{alg:backoff_computation}, lines $34-37$), whereby the CW is adapted as $\text{CW} = \text{CW}_0 \cdot 2^{\min(n, N_\text{max})}$, being $\text{CW}_0$ the initial backoff, $n$ the number of consecutive collisions, and $N_\text{max}$ the maximum CW stage.

\subsection{Deterministic Backoff (DB)}

DB aims at providing determinism to the channel access. For that, it proposes computing the backoff deterministically using a base backoff parameter, $b$, and the number of interruptions experienced during the last backoff decrease, $\text{IPT}$, so that $BO = b + \text{IPT}$ (Alg.~\ref{alg:backoff_computation}, line $38$). The deterministic backoff, as defined in~\cite{kosek2022db}, can be disabled when collisions occur. In particular, a deterministic backoff is computed when $(n \mod m) < \beta$, where $m$ and $\beta$ are parameters that regulate the periodicity in which a deterministic backoff can be used even with collisions. Otherwise, a random backoff is computed. In this paper, we simplify such a condition and reestablish the random backoff for any $n>0$ (Alg.~\ref{alg:backoff_computation}, lines $34-37$).

\subsection{It's Your Turn (IYT)}

IYT is a backoff computation method that builds on top of the DCF, thus being compatible with the current IEEE 802.11 standard. IYT's backoff computation is based on environmental awareness so that neighboring inter-BSS activity is considered for achieving an ordered access to the medium and, hence, achieving higher determinism. The operations done by IYT are as follows:
\begin{enumerate}
    \item \textbf{Neighboring activity detection:} Inter-BSS transmissions are overheard to identify the presence of overlapping BSSs (Alg.~\ref{alg:backoff_computation}, line 3). Neighboring activity is done in practice by inspecting the headers of the detected inter-BSS frames (e.g., data frames, Beacon frames, Probe request/response), which include relevant information such as the source and destination address, or the BSS ID. Alternatively, the BSS Color, which is included in Physical Layer Convergence Protocol (PLCP) headers of 802.11ax and onwards frames, can be retrieved and used to quickly identify inter-BSS activity. In Alg.~\ref{alg:backoff_computation} (lines $10-12$), the neighbor list is updated upon detecting strong enough inter-BSS signals, so that the power received at the device of interest $j$, $\mathcal{P}(j)$ is above the CCA threshold.
    \item \textbf{Neighboring list ordering:} Based on the neighbor devices detected by $j$, an ordered list $\mathcal{L}_j$ is maintained, which determines the order in which channel accesses must be done. The list's order can be determined, for instance, based on BSS ID or the BSS Color of the different involved BSSs, ordered in ascending order. Using such an order, a policy like RR can be adopted. As a way of example, $\mathcal{L}$ = \{BSS$_2$, BSS$_1$, BSS$_3$\} would be obtained by any of the 3 BSSs using colors 0001, 0010, and 1011, respectively, for RR with ascending order. Notice that other ordering policies could be adopted, depending on traffic priorities, device conditions (e.g., devices with good channel conditions may access the channel more frequently), or any other aspects. 
    \item \textbf{Backoff computation:} The neighboring list is used to compute the backoff, $BO$, in a way that the established order $\mathcal{L}_j$ is respected. In particular, a token $T$ indicating the device to transmit next is maintained by each device $j$ to keep track of the status of the transmission order. The token is updated every time a device $i$ from the list ends a transmission (Alg.~\ref{alg:backoff_computation}, line 14), which allows computing the distance of the device of interest $j$ to the token (Alg.~\ref{alg:backoff_computation}, lines $15-18$). The distance to the token $d$ is used to compute the bounds [CW$_\text{min}-1$, CW$_\text{max}-1$] among which to sample the random backoff (Alg.~\ref{alg:backoff_computation}, lines $41-42$): $BO = U\sim[\text{CW}_\text{min}-1, \text{CW}_\text{max}-1]$, where CW$_\text{min}=d\cdot \text{CW}_0$ and CW$_\text{max}= (d+1)\cdot \text{CW}_0$.    
\end{enumerate}

The three mechanisms described above are further illustrated in Fig.~\ref{fig:transmission_diagrams}, which sketches the packet transmissions performed by two overlapping STAs, namely STA$1$ and STA$2$, in each case. First (Fig.~\ref{fig:beb}), since BEB implements a random BO computation, multiple consecutive TXOPs can be held by STA$2$, to the detriment of STA$1$'s latency. Second (Fig.~\ref{fig:db}),  DB leads to a deterministic sequence where the two STAs alternate their transmissions. Finally (Fig.~\ref{fig:iyt}), IYT achieves the same ordering effect as DB but through a different approach, which is based on creating awareness about neighboring activity. As we show in the next section, the approach taken by IYT leads to higher flexibility and greater benefits than for DB.

\section{Performance Evaluation}
\label{sec:performance_evaluation}

In this section, we evaluate IYT through simulations and compare its performance and properties against BEB and DB. The simulations were done using Komondor~\cite{barrachina2019komondor} with the main configuration parameters collected in Table~\ref{tbl:simulation_parameters}. For further details about the considered frame types, their sizes, and inter-frame periods, refer to~\cite[Table~B.6]{wilhelmi2021spatial}.

\begin{table}[ht!]
\centering
\caption{Simulation parameters.}
\label{tbl:simulation_parameters}
\resizebox{.8\columnwidth}{!}{%
\begin{tabular}{@{}clc@{}}
\toprule
Parameter & \multicolumn{1}{c}{\textbf{Description}} & \textbf{Value} \\ \midrule
$t$ & Simulation time & 100 s \\
$N_\text{sim}$ & Number of random deployments & 100 \\
$N_\text{BSS}$ & Number of BSSs & $1-9$ \\
$F_c$ & Carrier frequency & 6 GHz \\
$B$ & Transmission bandwidth & 20 MHz \\
$\mathcal{P}^\text{Noise}$ & Noise power & -95 dBm \\
$\mathcal{P}_{tx,\max}$ & Max. transmit power & 20 dBm \\
CCA & CCA threshold & -82 dBm \\
$S$ & Single-user spatial streams & 1 \\
$G^\text{TX/RX}$ & Transmitter/receiver antenna gain & 0/0 dBi \\
$\gamma_{\textrm{CE}}$ & Capture effect threshold & 10 dB \\
PL & Path loss model & See \cite{wilhelmi2021spatial} \\
$PL_0$ & Loss at the reference dist. & 5 dB \\
$\nu$ & Path-loss exponent & 4.4 \\
$\sigma$ & Shadowing factor & 9.5 dB\\
$\omega$ & Obstacles factor & 30 dB\\
$\text{TXOP}_{\max}$ & TXOP duration limit & 5.484 ms\\ 
$\text{A-MPDU}_{\max}$ & A-MPDU size & 64\\ 
$L_{D}$ & Length of data packets & 1500 bytes \\ 
$\mathcal{T}$ & Traffic model & Full-buffer \\ 
\midrule
$\text{CW}_0$ & Initial contention window & 16 \\
$N_{\max}$ & Max. CW stage & 5 \\
$b$ & DB's base backoff & 5 \\
\bottomrule
\end{tabular}%
}
\end{table}

\subsection{Behavior in a toy scenario}

\begin{figure}[t!]
    \centering
    \begin{subfigure}{.49\columnwidth}
        \centering
        \includegraphics[width=\linewidth]{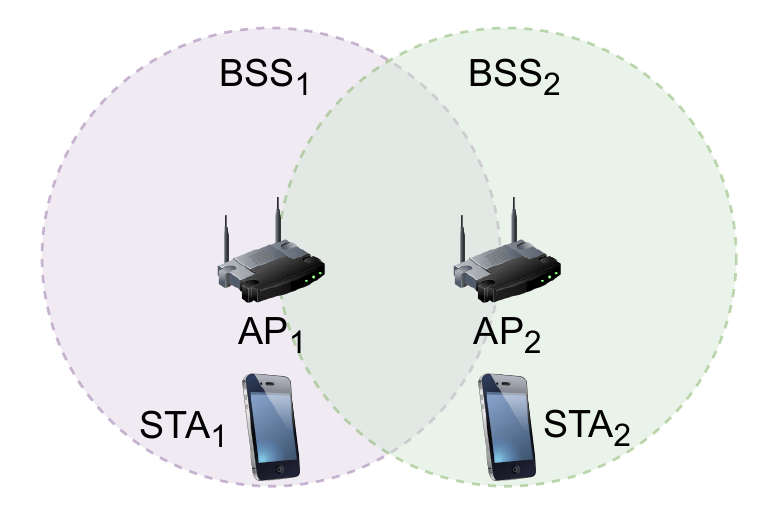}
        \caption{}
        \label{fig:toy_scenario_1a}
     \end{subfigure}
     \hfill
     \begin{subfigure}{.49\columnwidth}
        \centering
        \includegraphics[width=\linewidth]{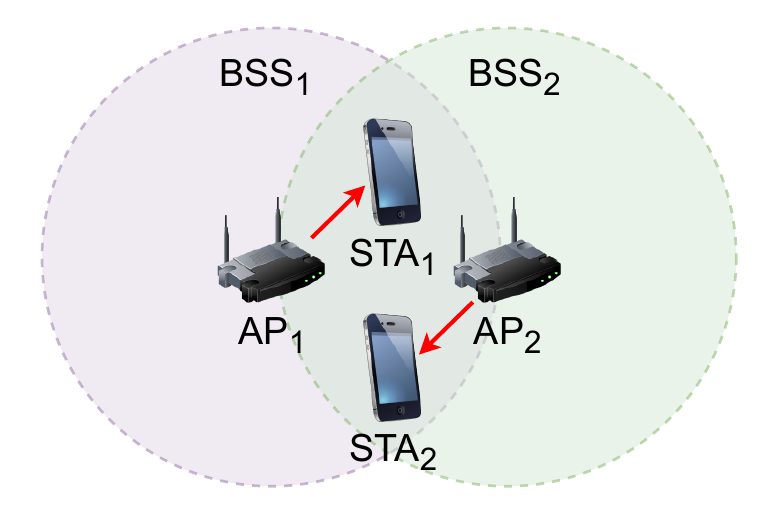}
        \caption{}
        \label{fig:toy_scenario_1b}
     \end{subfigure}
     \caption{Toy scenario formed by two BSSs. In (a), no packet losses occur when BSS 1 and BSS 2 transmit concurrently, while in (b), they do.}
    \label{fig:toy_scenario}
\end{figure}

\begin{figure}[t!]
    \centering
    \begin{subfigure}{.49\columnwidth}
        \centering
        \includegraphics[width=\linewidth]{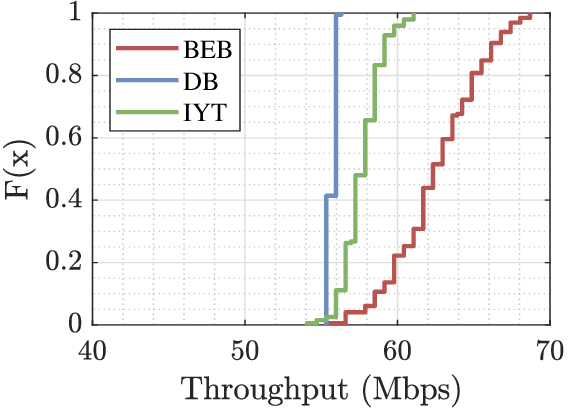}
        \caption{}
        \label{fig:cdf_throughput_toy_scenario_1a}
     \end{subfigure}
     \hfill
     \begin{subfigure}{.49\columnwidth}
        \centering
        \includegraphics[width=\linewidth]{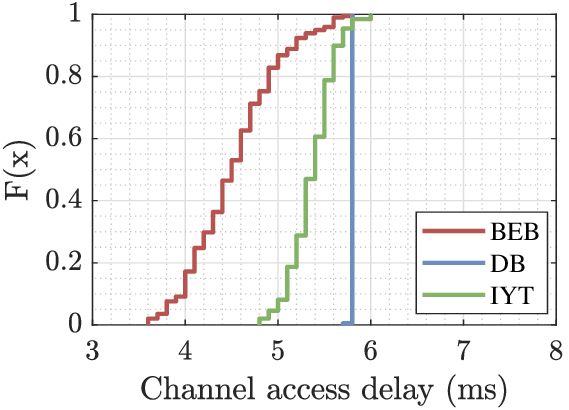}
        \caption{}
        \label{fig:cdf_access_delay_toy_scenario_1a}
     \end{subfigure}
     \begin{subfigure}{.49\columnwidth}
        \centering
        \includegraphics[width=\linewidth]{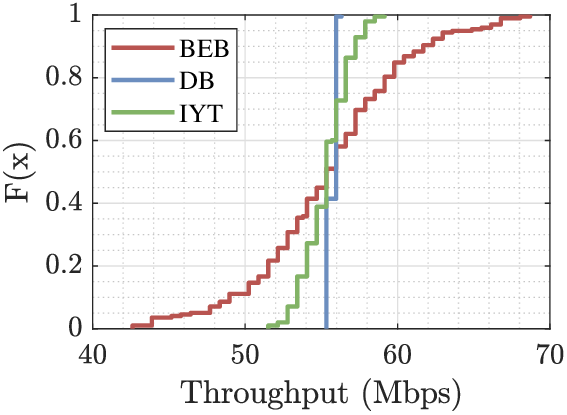}
        \caption{}
        \label{fig:cdf_throughput_toy_scenario_1b}
     \end{subfigure}
     \hfill
     \begin{subfigure}{.49\columnwidth}
        \centering
        \includegraphics[width=\linewidth]{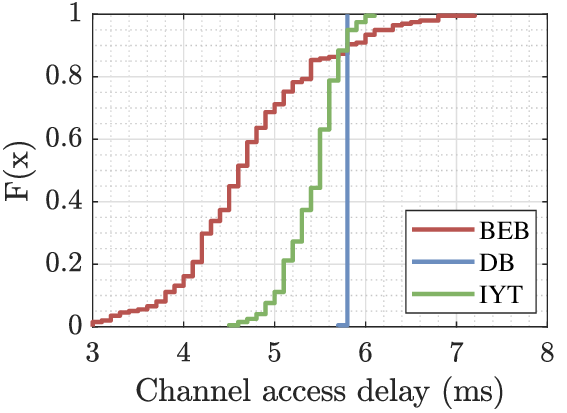}
        \caption{}
        \label{fig:cdf_access_delay_toy_scenario_1b}
     \end{subfigure}
     \caption{CDF of the performance achieved by BEB, DB, and IYT in the 2-BSS toy scenario. (a) Throughput (no losses), (b) channel access delay (no losses), (c) throughput (with losses), (d) channel access delay (with losses).}
    \label{fig:results_toy_scenario}
\end{figure}

We start our analysis focusing on the 2-BSS toy scenario illustrated in Fig.~\ref{fig:toy_scenario}, where the two considered BSSs (each composed of one AP and one STA) are overlapping. Within the same scenario, we consider two cases by selecting different STA locations, leading to a case where simultaneous transmissions (i.e., collisions) by the two APs do not lead to packet losses (Fig.~\ref{fig:toy_scenario_1a}) and another case where they do (Fig.~\ref{fig:toy_scenario_1b}). The throughput and the channel access delay achieved in each case are reported in Fig.~\ref{fig:results_toy_scenario}. The reported values include the mean values observed at each BSS every $\Delta=1$~s in a simulation of $t=100$~s duration. 

In the first case (no packet losses, Fig.~\ref{fig:cdf_throughput_toy_scenario_1a} and Fig.~\ref{fig:cdf_access_delay_toy_scenario_1a}), BEB provides the best but highly varying performance when compared to DB and IYT. The reason is that BEB leads to collisions by backoff when the two APs select the same random backoff, but that does not incur packet losses in this scenario, thus allowing to improve both throughput and channel access delay. DB and IYT, in contrast, provide an ordered channel access, so the transmissions from each BSS are alternated and no collisions (or very few) are experienced. This is an appealing property when packet losses occur (Fig.~\ref{fig:cdf_throughput_toy_scenario_1b} and Fig.~\ref{fig:cdf_access_delay_toy_scenario_1b}). In that case, DB and IYT provide a much more stable and deterministic performance than BEB, thus improving the worst-case performance. While DB provides very high determinism, IYT exhibits some variability but still provides good reliability and, more importantly, offers high flexibility by preserving random access properties. As we show next, IYT's flexibility is key for scalability and coexistence.

\subsection{Coexistence}

\begin{figure}[t!]
    \centering
    \begin{subfigure}{0.49\columnwidth}
        \centering
        \includegraphics[width=\linewidth]{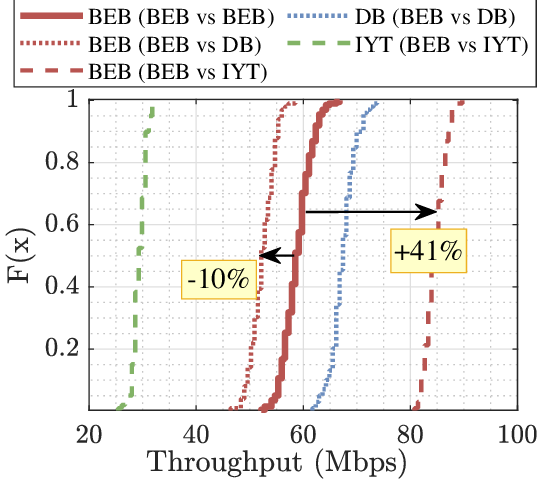}
        \caption{}
        \label{fig:results_coexistence_1a}
     \end{subfigure}
     \hfill
    \begin{subfigure}{0.49\columnwidth}
        \centering
        \includegraphics[width=\linewidth]{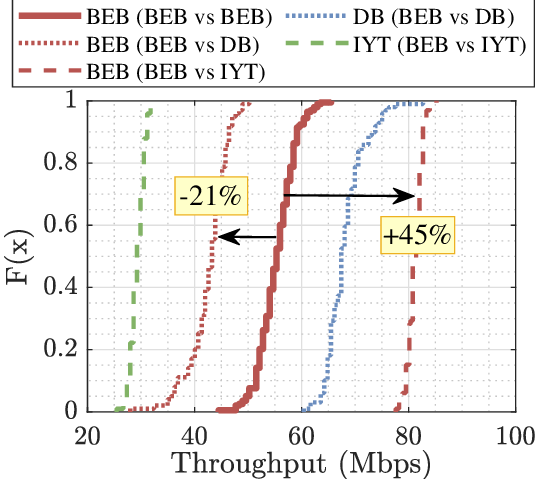}
        \caption{}
        \label{fig:results_coexistence_1b}
     \end{subfigure}
     \caption{CDF of the throughput achieved with different combinations of coexisting mechanisms in the 2-BSS toy scenario. (a) No packet losses, (b) with packet losses. Each line represents the performance of a device implementing a given mechanism (BEB, DB, or IYT) when competing with another device using another mechanism (BEB, DB, or IYT). The throughput values of both BSSs are aggregated in the same line for the legacy case (BEB vs BEB).}
    \label{fig:results_coexistence}
\end{figure}

Next, we show the capabilities of each mechanism when coexisting with legacy BEB devices. For that, we use the toy scenario from Fig.~\ref{fig:toy_scenario} and evaluate in Fig.~\ref{fig:results_coexistence} the performance of the two BSSs when one implements BEB and the other, each of the studied mechanisms (BEB, DB, or IYT). 

Without packet losses (see Fig.~\ref{fig:results_coexistence_1a}), the BEB device suffers a performance degradation of over 10\% when another one employs DB. This effect, as shown in Fig.~\ref{fig:results_coexistence_1b}, is increased with packet losses, so BEB's performance decreases by 21\%. In this particular case, the exponential backoff phases performed by BEB allow the DB device to access the channel more frequently. Contrary to the performance degradation incurred by DB, we observe that the BEB device achieves substantially better performance (over 41\% and 45\%, without and with packet losses, respectively) in the presence of an IYT device. This is, however, at the expense of the performance experienced by the IYT device, which turns out to be too courteous with the legacy BEB one. 

\begin{figure}[t!]
    \centering
    \includegraphics[width=.9\linewidth]{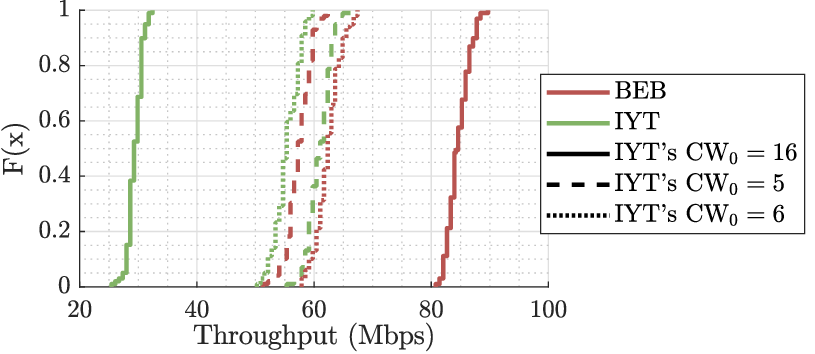}
    \caption{CDF of the throughput achieved by BEB and IYT devices in the 2-BSS toy scenario (no packet losses), for CW$_0\in\{5,6,16\}$ (CW$_0$ is always fixed to 16 for BEB). Each line represents the performance of a device implementing BEB (in red) when competing with another IYT device (in green).}      \label{fig:cdf_throughput_vs_token_improved_toy_scenario_3a}
\end{figure}

The low performance obtained by IYT is due to the fact it uses the same CW parameters as BEB. While IYT lets BEB access the channel when it is its turn, BEB is unaware of IYT, which entails the latter enjoying fewer opportunities to access the channel. Nevertheless, higher fairness can be achieved by adjusting the channel access parameters of IYT. This is supported by the results displayed in Fig.~\ref{fig:cdf_throughput_vs_token_improved_toy_scenario_3a}, where different CW values (CW$_0 = \{5, 6, 16\}$) for IYT are studied for the coexistence between BEB and IYT devices. As shown, tuning the CW for the IYT device leads to a fairer performance between BEB and IYT, with CW$_0 = 5$ being the value providing the most similar performance between the two BSSs.

\subsection{Scalability}

\begin{figure*}
    \centering
    \begin{subfigure}{0.6\columnwidth}
        \centering
        \includegraphics[width=\linewidth]{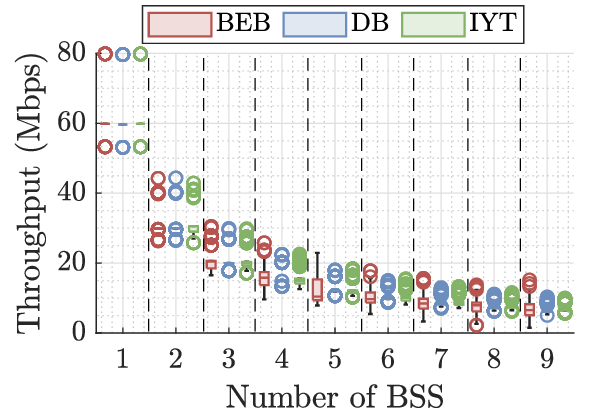}
        \caption{}
        \label{fig:throughput_overlapping_scenario}
     \end{subfigure}
     \hfill
     \begin{subfigure}{0.6\columnwidth}
        \centering
        \includegraphics[width=\linewidth]{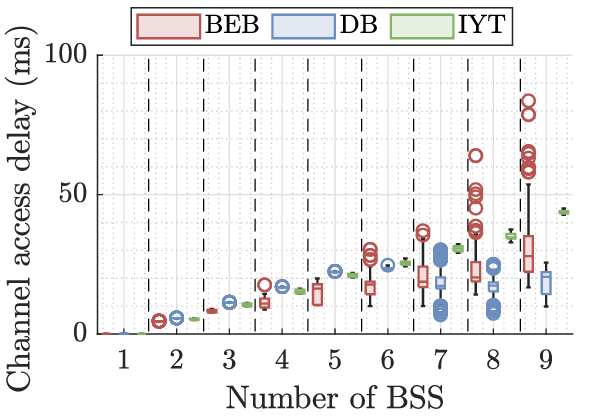}
        \caption{}
        \label{fig:delay_overlapping_scenario}
     \end{subfigure}
     \hfill
     \begin{subfigure}{0.6\columnwidth}
        \centering
        \includegraphics[width=\linewidth]{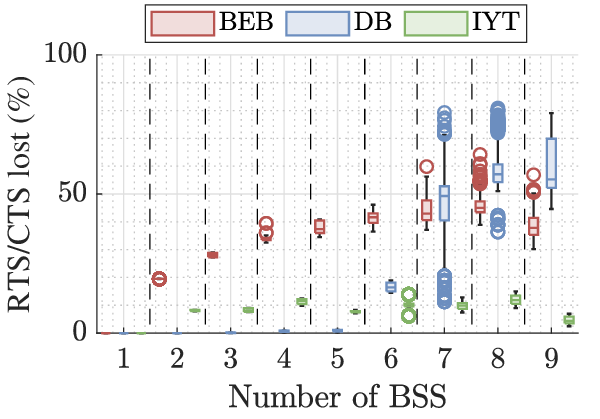}
        \caption{}
        \label{fig:collisions_overlapping_scenario}
     \end{subfigure}
     \caption{Performance achieved by BEB, DB, and IYT versus the number of overlapping BSSs. Each boxplot shows the median (line inside the box), the worst/best 25\% performance quartiles (top/bottom edges of the box), and the maximum/minimum values (whisker lines outside the box). (a) Throughput, (b) channel access delay, (c) RTS/CTS loss percentage.}
    \label{fig:results_overlapping_scenario}
\end{figure*}

Next, we study the scalability of the different mechanisms against the number of overlapping BSSs. For that, we consider different scenarios with $N_\text{BSS} \in [1,9]$ fully overlapping BSSs, where STAs are located at a random position between 3 and 4 meters away from their AP. The results obtained in this scenario, aggregating the outputs from $N_\text{sim}=100$ random deployments, are shown in Fig.~\ref{fig:results_overlapping_scenario}. 

As shown in Fig.~\ref{fig:throughput_overlapping_scenario}, the three mechanisms obtain very similar throughput for a low number of overlapping BSSs. As the number of BSSs increases, the behaviors of the different mechanisms diverge, with DB and IYT providing lower variability than BEB. When looking at the channel access delay (Fig.~\ref{fig:delay_overlapping_scenario}), we see that BEB leads to the lowest channel access delay for low BSS densities. However, BEB's performance degrades when density increases, thus resulting in poor scalability. And not only that, but the worst-case latency associated with BEB becomes particularly high when the number of overlapping devices is high. The reason behind those excessive delays lies in the number of collisions experienced by BEB (see Fig.~\ref{fig:collisions_overlapping_scenario}), which activate the exponential phases of backoff computation. Regarding DB, we observe a little variation within channel access delay values from 1 to 6 overlapping BSSs. However, from 7 BSSs, the channel access delay of DB becomes unstable and leads to highly varying delay values, which are motivated by the lack of adaptability of DB (parameter $b$ must be carefully configured as a function of the number of BSSs). Finally, IYT offers a slower but stable channel access delay across all the BSSs, which is almost deterministic for any number of overlapping BSSs (Fig.~\ref{fig:delay_overlapping_scenario}). The reason behind this is that, unlike BEB and DB, IYT keeps the number of collisions low (see Fig.~\ref{fig:collisions_overlapping_scenario}). In this regard, we identify a trade-off between the aggressiveness at which the channel is accessed, the collisions derived from that, and the resulting performance. While BEB and especially DB provide faster channel access than IYT, the latter provides higher reliability and fairer performance among all the involved BSSs.

\subsection{Overall performance in random deployments}

\begin{figure}[ht!]
    \centering
    \begin{subfigure}{0.47\columnwidth}
        \centering
        \includegraphics[width=\linewidth]{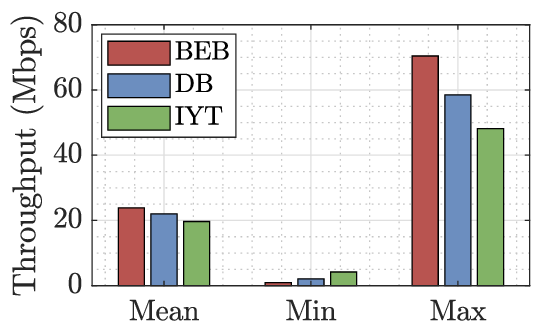}
        \caption{}
        \label{fig:barplot_throughput}
     \end{subfigure}
     \hfill
     \begin{subfigure}{0.47\columnwidth}
        \centering
        \includegraphics[width=\linewidth]{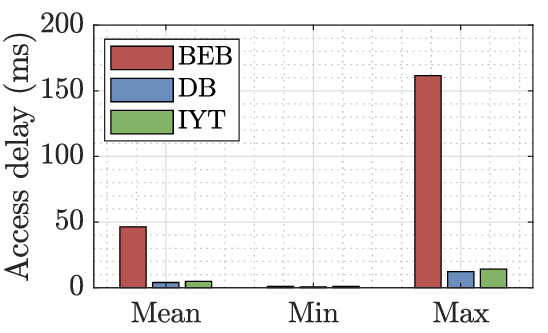}
        \caption{}
        \label{fig:barplot_delay}
     \end{subfigure}
     \hfill
     %
     %
     \caption{Average performance (mean, min, and max) achieved by BEB, DB, and IYT across $N_\text{sim} = 100$ random deployments. (a) Throughput, (b) channel access delay.}
    \label{fig:results_random_scenario}
\end{figure}

We now focus on the overall performance achieved by $N_\text{BSS} = 9$ BSSs (each with an AP located at the center of the cell and a randomly located STA) in a $3\times 3$ grid of $15\times 15$ meters and with frequency reuse $f = 3$. The average results from $N_\text{sim}=100$ random deployments are depicted in Fig.~\ref{fig:results_random_scenario}.

As shown in Fig.~\ref{fig:barplot_throughput}, BEB leads to the highest mean and maximum throughput but, in turn, to the lowest minimum. This is translated into the very high worst-case latency shown in Fig.~\ref{fig:barplot_throughput}, where BEB leads to up to $161$~ms. DB and IYT, in contrast, sacrifice the best performance across the deployment to improve the worst cases. In terms of minimum (worst-case) throughput (Fig.~\ref{fig:barplot_throughput}), this effect does not translate into very high gains, but, interestingly, it does so for the access delay (Fig.~\ref{fig:barplot_delay}), where the maximum (worst-case) latency is kept under control (31~ms and 25~ms for DB and IYT, respectively). As a result, the low throughput values experienced by DB and IYT are mostly associated with poor signal quality (STA are located far from the AP or exposed to too much interference), but, unlike for BEB, not to channel access competition.

\section{Conclusions}
\label{sec:conclusions}

In this paper, we proposed IYT, a novel backoff mechanism that improves the reliability of decentralized channel access in Wi-Fi. We evaluated IYT using the Komondor simulator and compared its performance against BEB and DB under full-buffer traffic conditions. Our results showed that IYT provides high determinism and adapts well to various situations, thus improving the worst-case performance compared to the other studied baselines (BEB and DB) and positioning itself as a strong candidate for improving reliability in Wi-Fi 8. We also showed some appealing properties of IYT regarding coexistence (it is legacy-friendly), scalability (it preserves determinism even when network density increases), and adaptability (it performs well in large, random deployments).

\bibliographystyle{IEEEtran}
\bibliography{bib}

\begin{thebibliography}{10}
\providecommand{\url}[1]{#1}
\csname url@samestyle\endcsname
\providecommand{\newblock}{\relax}
\providecommand{\bibinfo}[2]{#2}
\providecommand{\BIBentrySTDinterwordspacing}{\spaceskip=0pt\relax}
\providecommand{\BIBentryALTinterwordstretchfactor}{4}
\providecommand{\BIBentryALTinterwordspacing}{\spaceskip=\fontdimen2\font plus
\BIBentryALTinterwordstretchfactor\fontdimen3\font minus
  \fontdimen4\font\relax}
\providecommand{\BIBforeignlanguage}[2]{{%
\expandafter\ifx\csname l@#1\endcsname\relax
\typeout{** WARNING: IEEEtran.bst: No hyphenation pattern has been}%
\typeout{** loaded for the language `#1'. Using the pattern for}%
\typeout{** the default language instead.}%
\else
\language=\csname l@#1\endcsname
\fi
#2}}
\providecommand{\BIBdecl}{\relax}
\BIBdecl

\bibitem{GalGerCar2023}
{L. Galati-Giordano} \emph{et~al.}, ``{What Will Wi-Fi 8 Be? A Primer on IEEE
  802.11bn Ultra High Reliability},'' \emph{arXiv:2303.10442}, 2023.

\bibitem{abinader2014performance}
F.~M. Abinader \emph{et~al.}, ``{Performance evaluation of IEEE 802.11n WLAN in
  dense deployment scenarios},'' in \emph{2014 IEEE 80th vehicular technology
  conference (VTC2014-Fall)}.\hskip 1em plus 0.5em minus 0.4em\relax IEEE,
  2014, pp. 1--5.

\bibitem{carrascosa2024performance}
M.~Carrascosa-Zamacois \emph{et~al.}, ``{Performance Evaluation of MLO for XR
  Streaming: Can Wi-Fi 7 Meet the Expectations?}'' \emph{arXiv preprint
  arXiv:2407.05802}, 2024.

\bibitem{kosek2022db}
K.~Kosek-Szott \emph{et~al.}, ``{DB-LBT: Deterministic Backoff with Listen
  Before Talk for Wi-Fi/NR-U Coexistence in Shared Bands},'' in \emph{2022 30th
  International Symposium on Modeling, Analysis, and Simulation of Computer and
  Telecommunication Systems (MASCOTS)}.\hskip 1em plus 0.5em minus 0.4em\relax
  IEEE, 2022, pp. 168--175.

\bibitem{barcelo2009learning}
J.~Barcelo \emph{et~al.}, ``{Learning-BEB: avoiding collisions in WLANs},''
  \emph{Carrier Sense Multiple Access with Enhanced Collision Avoidance},
  p.~23, 2009.

\bibitem{tinnirello2009rethinking}
I.~Tinnirello and G.~Bianchi, ``{Rethinking the IEEE 802.11 e EDCA performance
  modeling methodology},'' \emph{IEEE/ACM transactions on networking}, vol.~18,
  no.~2, pp. 540--553, 2009.

\bibitem{hong2012channel}
K.~Hong \emph{et~al.}, ``{Channel condition based contention window adaptation
  in IEEE 802.11 WLANs},'' \emph{IEEE Transactions on Communications}, vol.~60,
  no.~2, pp. 469--478, 2012.

\bibitem{gawlowicz2021distributed}
P.~Gaw{\l}owicz \emph{et~al.}, ``{Distributed Learning for Proportional-Fair
  Resource Allocation in Coexisting WiFi Networks},'' in \emph{2021 19th
  International Symposium on Modeling and Optimization in Mobile, Ad hoc, and
  Wireless Networks (WiOpt)}.\hskip 1em plus 0.5em minus 0.4em\relax IEEE,
  2021, pp. 1--8.

\bibitem{rhee2005z}
I.~Rhee \emph{et~al.}, ``{Z-MAC: a hybrid MAC for wireless sensor networks},''
  in \emph{Proceedings of the 3rd international conference on Embedded
  networked sensor systems}, 2005, pp. 90--101.

\bibitem{patras2016rigorous}
P.~Patras \emph{et~al.}, ``{Rigorous and practical proportional-fair allocation
  for multi-rate Wi-Fi},'' \emph{Ad Hoc Networks}, vol.~36, pp. 21--34, 2016.

\bibitem{kim2017centralized}
J.~D. Kim \emph{et~al.}, ``{Centralized random backoff for collision resolution
  in Wi-Fi networks},'' \emph{IEEE Transactions on Wireless Communications},
  vol.~16, no.~9, pp. 5838--5852, 2017.

\bibitem{huawei2017coexistence}
H.~Huawei, ``{Coexistence and channel access for NR-based unlicensed band
  operation},'' in \emph{R1-1711467, 3GPP TSG RAN WG1 NR Ad Hoc Meeting,
  Qingdao, China, dated Jun}, 2017, pp. 27--30.

\bibitem{lagen2019new}
S.~Lagen \emph{et~al.}, ``{New radio beam-based access to unlicensed spectrum:
  Design challenges and solutions},'' \emph{IEEE Communications Surveys \&
  Tutorials}, vol.~22, no.~1, pp. 8--37, 2019.

\bibitem{barrachina2019komondor}
S.~Barrachina-Munoz \emph{et~al.}, ``{Komondor: A wireless network simulator
  for next-generation high-density WLANs},'' in \emph{2019 Wireless Days
  (WD)}.\hskip 1em plus 0.5em minus 0.4em\relax IEEE, 2019, pp. 1--8.

\bibitem{wilhelmi2021spatial}
F.~Wilhelmi \emph{et~al.}, ``{Spatial reuse in IEEE 802.11ax WLANs},''
  \emph{Computer Communications}, vol. 170, pp. 65--83, 2021.

\end{thebibliography}

\end{document}